\documentclass[conference]{IEEEtran}
\IEEEoverridecommandlockouts
\usepackage{cite}
\usepackage{amsmath,amssymb,amsfonts}
\usepackage{algorithmic}
\usepackage{graphicx}
\usepackage{textcomp}
\usepackage{xcolor}
\usepackage{booktabs}
\begin{document}

\title{Information Reconciliation for High-Dimensional Quantum Key Distribution using Nonbinary LDPC codes\\

\thanks{We acknowledge funding from: The Center of Excellence SPOC (ref DNRF123)}
}


\author{\IEEEauthorblockN{Ronny M\"uller\IEEEauthorrefmark{2}, Davide Bacco\IEEEauthorrefmark{1}, Leif Katsou Oxenløwe\IEEEauthorrefmark{2}, and Søren Forchhammer\IEEEauthorrefmark{2}}
\IEEEauthorblockA{\IEEEauthorrefmark{2}Department of Electrical and Photonics Engineering\\
Technical University of Denmark, Lyngby, Denmark\\
Email: ronmu@dtu.dk}
\IEEEauthorblockA{\IEEEauthorrefmark{1}Department of Physics and Astronomy\\ University of Florence, Florence, Italy}}

\maketitle

\begin{abstract}
Information Reconciliation is an essential part of Quantum Key distribution protocols that closely resembles  Slepian-Wolf coding. The application of nonbinary LDPC codes in the Information Reconciliation stage of a high-dimensional discrete-variable Quantum Key Distribution setup is proposed. We model the quantum channel using a $q$-ary symmetric channel over which qudits are sent. Node degree distributions optimized via density evolution for the Quantum Key Distribution setting are presented, and we show that codes constructed using these distributions allow for efficient reconciliation of large-alphabet keys.
\end{abstract}


\section{Introduction}
Quantum Key Distribution (QKD) aims to share a symmetric secret key usable for encryption between two remote parties. It consists of a quantum phase in which the two parties (Alice and Bob) share some quantum information sent over a quantum channel, followed by a classical post-processing phase in which a shared secret key is extracted \cite{Bennett_2014}. While QKD using quantum bits (qubits) is well established, its High-dimensional version (HD-QKD) (using qu\textit{d}its) has seen significantly less research effort. HD-QKD offers many potential advantages, most importantly  a significantly increased resilience to noise \cite{PhysRevA.61.062308}. Physical implementations of HD-QKD systems have been shown using a variety of different technologies, including but not limited to time-bin encoding \cite{PhysRevApplied.14.014051}, time-energy encoding \cite{time-en}, and orbital angular momentum \cite{PhysRevApplied.11.064058}.\\
An essential part of the post-processing is the information reconciliation phase during which a common string is extracted from the results of the quantum state measurements. This is followed by privacy amplification \cite{pa}, during which a secret key is extracted from that common string, e.g. using universal hash functions.  While a lot of research has analyzed and optimized the information reconciliation phase for the two-dimensional case, little regard has yet been paid to the high-dimensional case,  apart from the introduction of the layered scheme in 2013 \cite{6502993}. In contrast to most channel coding applications, the (HD)-QKD setting has lower requirements on latency and throughput but puts a lot of value on a minimized information leakage. Motivated by this setting, the excellent decoding performance of nonbinary LDPC codes \cite{706440}, and their native handling of high dimensions, we explore the design and application of nonbinary LDPC codes for the post-processing in HD-QKD protocols.

\section{Background}
\subsection{Information Reconciliation}

The information reconciliation stage in QKD aims to correct any mismatch between the keys of the two parties while leaking as little information as possible to a potential eavesdropper. In general, Alice sends a random string $\mathbf{x}=(x_0,...,x_{n-1})$, $x_i = \{0,...,q-1\}$ of $n$ qudits of dim $q$ to Bob who proceeds to measure them, resulting in his version of the string $\mathbf{y}=(y_0,...,y_{n-1})$, $y_i = \{0,...,q-1\}$. We assume that the quantum channel can be accurately modeled by a substitute channel where $\mathbf{x}$ and $\mathbf{y}$ are correlated as a $q$-ary symmetric channel, as errors are usually uncorrelated and symmetric. The transition probabilities of such a channel is given by:

\begin{equation}
    \text{P}(y_i|x_i) = \begin{cases} 1-p & y_i=x_i,\\
    \frac{p}{1-q} & \text{else}.
    \end{cases}
\end{equation}
Here, the parameter $p$ denotes the channel transition probability. We refer to the symbol error rate between $\mathbf{x}$ and $\mathbf{y}$ as the quantum bit error rate (QBER) in a slight abuse of notation but in accordance with experimental works on HD-QKD. We assume the QBER to be an underlying channel property in our simulations, making it equivalent to the channel parameter $p$. Besides the qudits, Alice additionally sends a message $\mathbf{s}$ (usually the syndrome) of length $m$ over a classical communication channel which is assumed to be error-free. From a coding point of view, this is equal to  asymmetric Slepian-Wolf coding with side information at the receiver, where $\mathbf{s}$ represents the compressed version of $\mathbf{x}$ and $\mathbf{y}$ is the side information. A more detailed description of this equivalence can be found in \cite{1042242}. Any information that is leaked to a potential eavesdropper at any point of the quantum key distribution has to be subtracted from the final secret key \cite{Br_dler_2016} during privacy amplification. The information leaked during the information reconciliation stage will be denoted by leak$_{\text{IR}}$. Given no rate adaption, it can be upper-bounded by the syndrome length in symbols, $\text{leak}_{\text{IR}} \leq m$. Using the Slepian-Wolf bound \cite{1055037}, the minimum amount of leaked information to successfully reconcile with arbitrary low failure chance is therefore given by the conditional entropy:

\begin{equation}
    m  \geq n\text{H}(X|Y).
\end{equation}

The conditional entropy of the $q$-ary symmetric channel, assuming  independent and identically distributed input $X$, can be written as
\begin{equation}
    \text{H}(X|Y) = -((1-p)\text{log}_q(1-p) - p\cdot \text{log}_q(\frac{p}{q-1})).
\end{equation}

The performance of a code with respect to information leakage can be measured by its efficiency $f$, given by 

\begin{equation}
    f = \frac{m}{n\text{H}(X|Y)}.
\end{equation}

Note that an efficiency of $f>1$ corresponds to leaking more bits than required by the theoretical minimum of $f=1$, which represents the best possible performance according to the Slepian-Wolf bound. Practical systems have $f>1$.

\begin{table*}[h] 
\centering 

\caption{The code rate, the best possible theoretical threshold (TT) according to the Slepian-Wolf bound, the ensemble threshold (ET) using density evolution, the ensemble efficiency (EE), the evaluated code efficiency (CE), and the respective degree polynomial for an 8-ary symmetric channel.} 
\begin{tabular}{l c c c c l} 
Rate & TT & ET & EE & CE & Ensemble (edge view)\\ 
\midrule 

0.90 & 0.033 & 0.031 & 1.060 & 1.17 &
$\lambda(x) = 0.112x + 0.103x^2 + 0.194x^3 + 0.146x^9 + 0.163x^{10} + 0.003x^{17}$\\
&&&&&\hspace{0.9cm} $+ 0.173x^{19} + 0.049x^{26} + 0.006x^{28} + 0.052x^{29}$\\[2mm]

0.85 & 0.053 & 0.052  & 1.080 & 1.12 &$\lambda(x) = 0.125x + 0.165x^2 + 0.163x^5 + 0.11x^7 + 0.073x^{12} + 0.089x^{18}$\\
&&&&&\hspace{0.9cm} $+ 0.122x^{27} + 0.154x^{32}$\\[2mm]

0.80 & 0.0758 & 0.0724 & 1.038 & 1.12 & $\lambda(x) = 0.146x + 0.177x^2 + 0.130x^4 + 0.084x^7 + 0.149x^{10} + 0.035x^{19}$\\
&&&&&\hspace{0.9cm} $+ 0.029x^{22} + 0.087x^{25} + 0.163x^{26}$\\[2mm]

0.75 & 0.100 & 0.096 & 1.030 & 1.10 & $\lambda(x) = 0.165x + 0.192x^2 + 0.092x^5 + 0.176x^7 + 0.019x^{10} + 0.086x^{17}$\\
&&&&&\hspace{0.9cm} $+ 0.129x^{19} + 0.103x^{30} + 0.038x^{31}$\\[2mm]

0.70 & 0.126 & 0.121 & 1.030 & 1.10 & $\lambda(x) = 0.160x + 0.208x^2 + 0.140x^5 + 0.096x^8 + 0.028x^{10} + 0.013x^{11}$\\
&&&&&\hspace{0.9cm} $+ 0.113x^{18} + 0.032x^{21} + 0.211x^{27}$\\[2mm]

0.65 & 0.154 & 0.147 & 1.032 & 1.10 & $\lambda(x) = 0.173x + 0.228x^2 + 0.092x^4 + 0.169x^8 + 0.112x^{14} + 0.019x^{23}$\\
&&&&&\hspace{0.9cm} $+ 0.012x^{24} + 0.195x^{28}$\\[2mm]

0.60 & 0.183 & 0.177 & 1.026 & 1.10 & $\lambda(x) = 0.192x + 0.196x^2 + 0.222x^5 + 0.104x^{13} + 0.114x^{23} + 0.055x^{25}$\\
&&&&&\hspace{0.9cm} $+ 0.117x^{27}$\\[2mm]

0.55 & 0.214 & 0.207 & 1.024 & 1.10 & $\lambda(x) = 0.183x + 0.269x^2 + 0.124x^6 + 0.036x^8 + 0.097x^{10} + 0.004^{21}$\\
&&&&&\hspace{0.9cm} $+ 0.116x^{25} + 0.171x^{26}$\\[2mm]

0.50 & 0.247 & 0.239 & 1.024 & 1.11 & $\lambda(x) = 0.215x + 0.256x^2 + 0.030x^4 + 0.154x^7 + 0.065x^{11} + 0.050^{13}$\\
&&&&&\hspace{0.9cm} $+ 0.072x^{21} + 0.128x^{27}$\\[2mm]

\end{tabular}

\label{tab:lamda} 
\end{table*}

\begin{figure*}[h]
    \centering
    \includegraphics[width=\linewidth]{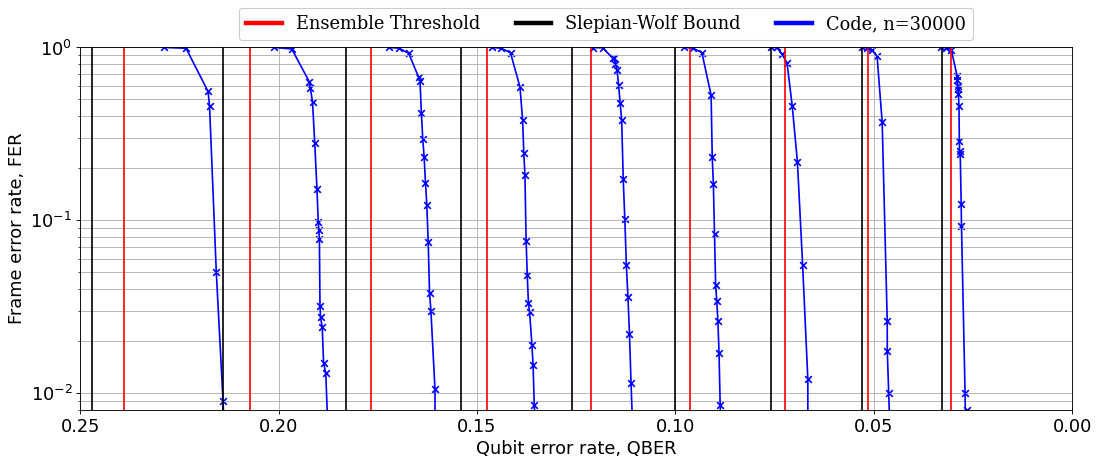}
    \caption{The frame error rate (FER) of 9 codes of coding rate [0.50, 0.55,...,0.90] (left to right) for different values of the QBER on a $q=8$-dimensional $q$-ary symmetric channel. Each data point has been evaluated with 5000 samples or until 100 errors have been collected.}
    \label{fig:fers}
\end{figure*}

\subsection{Nonbinary LDPC codes}

We assume some familiarity with belief propagation decoding of binary Low-density parity-check (LDPC) codes and Tanner graphs. For an overview see \cite{ldpcbook}.
Nonbinary LDPC codes can be characterized by their parity check Matrix $\mathbf{H}$, with $m$ rows and $n$ columns, which contain values in a Galois Field (GF) of order $q$ as its entries. To increase clarity in this section, we will mark all variables that represent a Galois field element with a hat, e.g.  $\hat{a}$. Furthermore, let $\oplus, \ominus, \otimes,$ and $\oslash$ refer to the common operations on Galois field elements.
An LDPC code can be represented as a bipartite graph, the Tanner graph. There, the parity-check equations constitute one side of the graph, called the check nodes, while the symbols of the codeword represent the other side of the graph, the variable nodes. The Tanner graph of a nonbinary LDPC code additionally has weighted edges between the check and variable nodes where the weight corresponds to the respective entry of $\mathbf{H}$. The syndrome $\mathbf{s}$ of the $q$-ary string $\mathbf{x}$ is calculated as $\mathbf{s} = \mathbf{H}\mathbf{x}$.

In this work, we deploy a log-domain FFT-SPA \cite{6577118, 1312606} for decoding. Detailed descriptions of this algorithm can be found in \cite{4787626, 6987278} but will be reiterated here for completeness of the work. Let $Z$ denote a random variable that takes values in GF($q$) such that $\text{P}(Z_i = k)$ denotes the probability that the qudit $i$ has value $k=0,...,q-1$. The probability vector $\mathbf{p}=(p_0,...p_{q-1})$, $p_j = \text{P}(Z=j)$ can be transformed into the log-domain using the generalized equivalent of the log-likelihood-ratio (LLR) in the binary case, $\mathbf{m}=(m_0,...,m_{q-1})$, $m_j = \text{log}\frac{\text{P}(Z=0)}{\text{P}(Z=j)} = \log(\frac{p_0}{p_j})$.  Given the LLR presentation, one can go back to probabilities via $p_j = \exp(-m_j)/\sum_{k=0}^{q-1} \exp(-m_k)$. We denote these transforms with $p(\cdot)$ and $m(\cdot)$. To further simplify notation, we define the multiplication and division of an element $\hat{a}$ in GF$(q)$ and an LLR message as a permutation of the indices of the vector:
\begin{align}
     \hat{a} \cdot \mathbf{m} &:= (m_{\hat{0} \oslash \hat{a}},...,m_{\hat{q-1} \oslash \hat{a}})\\
     \mathbf{m} / \hat{a}     &:= (m_{\hat{0}\otimes \hat{a}},...,m_{\hat{q-1} \otimes \hat{a}}),
\end{align}

where the multiplication and division of the indices take place in the Galois Field. These permutations are required as  we have to weigh messages according to their edge weight during the decoding. We further define two transformations involved with the decoding,
\begin{align}
    \Bar{\mathcal{F}}(\mathbf{m}, \hat{H}_{ij}) &= \mathcal{F}(p(\hat{H}_{ij}\cdot \mathbf{m}))\\
    \Bar{\mathcal{F}}(\mathbf{m}, \hat{H}_{ij})^{-1} &= m(\mathcal{F}^{-1}(\mathbf{m}))/\hat{H}_{ij},
\end{align}
where $\mathcal{F}$ is the discrete Fourier transform. Note that for $q$ being a power of 2, the Fast Walsh Hadamard Transform can be used. The decoding process then consists of two iterating message-passing phases, from check nodes to variable nodes and vice versa. The message update rule at iteration $l$ for the check node corresponding to the parity check matrix entry at $(i,j)$ can then be written as
\begin{equation}
    \mathbf{m}^{(l)}_{ij,\text{CV}} = \mathcal{A}(\hat{s}^{\prime}_i) \Bar{\mathcal{F}}^{-1}( \underset{j\prime \in \mathcal{M}(i)/j}{\Pi} \Bar{\mathcal{F}}(\mathbf{m}^{(l-1)}_{ij\prime}, \hat{H}_{ij\prime}), \hat{H}_{ij}),
\end{equation}

where $\mathcal{M}(i)$ is the set of all check nodes in row $i$ of $\mathbf{H}$. $\mathcal{A}$, defined as $\mathcal{A}_{kj}(\hat{a}) = \delta( \hat{a} \oplus k \ominus j) - \delta(a\ominus j)$, accounts for the nonzero syndrome \cite{6987278}. The weighted syndrome value is calculated as $\hat{s}^{\prime}_i = \hat{s}_i \oslash \hat{H}_{ij}$. The a posteriori message of column $j$ can be written as
\begin{equation}
    \Tilde{\mathbf{m}}^{(l)}_j = \mathbf{m}^{(0)}(j) + \underset{i^{\prime} \in \mathcal{N}(j)}{\sum} \mathbf{m}^{(l)}_{i^{\prime}j,\text{CV}},
\end{equation}
where $\mathcal{N}(j)$ is the set of all check nodes in column $j$ of $\mathbf{H}$. The best guess $\Tilde{\mathbf{x}}$ at each iteration $l$ can be calculated as the minimum value of the a posteriori, $\Tilde{x}_j^{(l)} = \text{argmin} (\Tilde{m}_j)$. The second message passings, from variable to check nodes, are given by
\begin{equation}
    \mathbf{m}_{ij, \text{VC}}^{(l)} = \Tilde{\mathbf{m}}^{(l)}_j - \mathbf{m}^{(l)}_{ij, \text{CV}}.
\end{equation}

The message passing is repeated until either $\mathbf{H}\Tilde{\mathbf{x}} = \mathbf{s}$ or the maximum number of iterations is reached.

\subsection{Density Evolution}

For a uniform edge weight distribution, the asymptotic decoding performance of LDPC codes in the regime of infinite code length is completely characterized by two polynomials \cite{4712621, 910577}:

\begin{equation}
    \lambda(x) = \sum_{i=0}^{d_{\text{v, max}}} \lambda_i x^{i-1} \quad \rho(x) = \sum_{i=0}^{d_{\text{c, max}}} \rho_i x^{i-1}.
\end{equation}

Here, $\lambda_i$ ($\rho_i$) are the fraction of edges connected to variable (check) nodes of degree $i$ while $d_{\text{v, max}}$ ($d_{\text{c, max}}$) is the highest degree of the variable (check) nodes. 
One can then define the code ensemble $\mathcal{E}(\lambda, \rho)$, representing all codes of infinite length with degree distributions given by $\lambda$ and $\rho$. The threshold $p_t(\lambda, \rho)$ of the code ensemble $\mathcal{E}(\lambda, \rho)$ is then defined as the worst channel parameter (QBER) for which decoding is still possible with an arbitrary small failure probability. This threshold can be approximated using Monte-Carlo Density Evolution (MC-DE), which is described in detail in \cite{inproceedings}. This method repeatedly samples the node degrees according to $\lambda$ and $\rho$, and draws random connections between nodes for each iteration. Given a high enough sample size, this simulates the performance of a cycle-free code. Note that MC-DE is especially well suited for nonbinary LDPC codes as the different edge weights help in decorrelating messages \cite{inproceedings}. During the simulation, we monitor the mean entropy of all messages. Once it drops below a certain value, the decoding is declared a success. If this does not happen after a maximum number of iterations, the evaluated channel parameter is above the threshold of $\mathcal{E}(\lambda,\rho)$. Using a concentrated check node distribution (which is favorable according to \cite{910580}) and a fixed code rate, we can further simplify to $\mathcal{E}(\lambda)$. One can then use the threshold as an objective function to optimize the code design, which is commonly done using the Differential Evolution algorithm \cite{de}.

\section{Results}

While the described approach is feasible for any dimension $q$, we chose to analyze for an 8-dimensional system as this is in the range of reported implementations \cite{PhysRevLett.127.110505}.
We designed 9 codes with code rates between 0.50 and 0.90, corresponding to a QBER range between 0 and 24.7\%. We used 100000 nodes with a maximum of 150 iterations for the MC-DE and swept the QBER in 20 steps in a short range below the best possible threshold. For the Differential Evolution, we used population sizes between 15 and 50, a differential weight of 0.85, and a crossover probability of 0.7. We enforced a sparsity of at most 10 different degrees in the polynomial and choose $d_{\text{v, max}}=40$. The sparsity allowed for reasonable optimization complexity. We observed that higher $d_{\text{v, max}}$ can lead to numerical instability during decoding. The results of the optimization can be found in Tab. \ref{tab:lamda}. For the optimization and evaluation, we used the all-zero codeword assumption, which holds for the given scenario of a symmetric channel  \cite{4787626}. For all rates, the designed thresholds are close to the theoretical bound. We constructed LDPC codes with a length of $n=30000$ symbols using Progressive Edge Growth \cite{1377521} to confirm the results of the density evolution experimentally. A log-FFT-SPA decoder is used to reconcile the messages. The performance of the finite-size codes can be seen in Fig. \ref{fig:fers}. As expected for finite-size codes, they do not reach the ensemble threshold but show sub-optimal performance. To estimate the impact on the secret key rate, we also evaluate the efficiency $f$ of the ensembles and the constructed codes at their working points. We measured the efficiency at a frame error rate (FER) of 1\% which is a common heuristic. Note that there is a direct trade-off between efficiency and FER which can be optimized to maximize the overall secret key generation. After the reconciliation, the correctness of the frames is asserted by exchanging short hashes. Incorrect frames are then simply discarded, only lowering the secret key rate. QKD systems can therefore efficiently operate at significantly higher FERs compared to what would be acceptable in a classical communication setting. The resulting efficiencies can be seen in Tab. \ref{tab:lamda}. The decoder runs a maximum of 100 decoding iterations.

\section{Discussion}

Nonbinary LDPC codes pose as a natural candidate for the information reconciliation stage of HD-QKD as their order can be matched to the dimension of the used qudits, and nonbinary codes are known to have good decoding performance \cite{706440}. Their main disadvantage, high decoding complexity, is less of an obstacle in this setting as the keys can be processed and stored before their usage in real-time applications, reducing the importance of decoding latency. We expect HD-QKD systems to be used in a long-haul setting with a high error rate and small count rate resulting in low throughput, further reducing the requirements on throughput volume. Nevertheless, less complex decoder algorithms like EMS \cite{4155118} or TEMS \cite{6125307} can be considered to allow the usage of longer codes. This is regarded beneficial as they reduce finite-size penalties of the secret key rate. Further, our codes allow for straightforward rate adaption and integration into existing schemes developed for binary channels, like Blind Reconciliation \cite{blind}.

While the efficiencies we report are between 1.10 and 1.17, and therefore in a similar range to the efficiencies of binary LDPC codes for conventional two-dimensional QKD \cite{Kiktenko_2017}, it is difficult to make statements about their performance in comparison to alternative schemes. To the best of our knowledge, only the layered scheme \cite{6502993}, which handles the decoding using $\lceil \log_2(q)\rceil $ binary decoders on bit planes, has been introduced for information reconciliation with regard to HD-QKD. It is similar in concept to the multilevel coding and multistage decoding methods used in slice reconciliation for continuous-variable (CV) QKD \cite{1266817}. While the layered scheme allows for reconciliation using binary LDPC codes only, it brings its own drawbacks, like error propagation, bit mapping, and interactive communication. Unfortunately, little to no results that would allow for a fair comparison have been published yet. Simulation results for $q=32$, a code length per layer of 800 symbols, and a QBER of 20\%, show an efficiency of the layered scheme of $f\approx1.4$ (Fig. 5 in \cite{6502993}). Later experimental implementations report efficiencies of 1.25 \cite{latest} ($q=3$, $n=1944$, $p=8\%$) and 1.17 \cite{2015NJPh...17b2002Z} ($q=1024$, $n=4000$, $p=39.6\%)$. These papers report their efficiencies in the $\beta$-notation. $\beta$ is commonly used in the continuous-variable QKD community, whereas $f$ is more widespread with respect to discrete-variable QKD. They can be related via $\beta (\text{H}(X)-\text{H}(X|Y)) = \text{H}(X)-f\text{H}(X|Y)$. Note that this conversion is dependent on $p$ and $q$. We want to emphasize that a direct comparison to the efficiencies reported in our work cannot be considered fair, as the settings are not identical. A fair comparison between the two approaches that additionally considers practicality remains the object of future work.

While the found ensembles show thresholds close to the Slepian-Wolf bound, we believe that even better results could be achieved by an extended search of the hyperparameters involved in the optimization, e.g. the enforced sparsity and highest degree of $\lambda$, and a finer sweep of the QBER during density evolution. Similarly, we believe that the performance of the constructed codes can be increased by using longer codes and improved versions of the PEG algorithm \cite{IPEG1, IPEG2}.

\section{Conclusion}

In this paper, information reconciliation for High-Dimensional Quantum Key Distribution using nonbinary LDPC codes over the $q$-ary symmetric channel is introduced. We present optimized node degree distributions for the case of $q=8$ that reach close to the Slepian-Wolf bound. Codes constructed according to these distributions with a length of 30000 symbols reach efficiencies of $f \approx 1.10$.



\bibliography{conf}

\end{document}